\documentclass[aps, prd, nofootinbib, reprint, preprintnumbers, superscriptaddress, showkeys]{revtex4-2}

\usepackage{amsmath}
\usepackage{amssymb}
\usepackage{tabularx}
\usepackage{graphicx}
\usepackage{float}
\usepackage[usenames,dvipsnames,svgnames]{xcolor}
\usepackage{hyperref}
\usepackage{nameref}
\usepackage[T1]{fontenc}

\hypersetup{dvips,dvipdfm,colorlinks=true,urlcolor=black,filecolor=magenta,linktoc=page,citecolor=red,linkcolor=blue,bookmarks=true}

\begin{document}

\title{The peculiar case of the Viaggiu holographic dark energy}

\author{Somnath Saha}
\email{sahasomnath847@gmail.com}
\affiliation{Department of Mathematics, Sree Chaitanya College, Habra 743268, West Bengal, India}

\author{Subhajit Saha}
\email{subhajit1729@gmail.com (Corresponding Author)}
\affiliation{Department of Mathematics, Panihati Mahavidyalaya, Kolkata 700110, West Bengal, India}

\author{Nilanjana Mahata}
\email{nilanjana_mahata@yahoo.com}
\affiliation{Department of Mathematics, Jadavpur University, Kolkata 700032, West Bengal, India}

%%%%%%%%%%%%%%%%%%%%%%%%%%%%%%%%%%%%%%%%%%%%%%%%%%%%%%%%%%%%%%%%%%%%%%%%%%%%%%%%%%%%%%%%%%%%%%%%%%%%%%%%%%%%%%%%%%%%%%%%%%%%%%%%%%%%%%%%%%%%%%%%%%%%%%%%%%%%%%

\begin{abstract}

\begin{center}
(Dated: The $11^{\text{th}}$ February, $2026$)
\end{center}

We study the plausibility of a holographic dark energy (HDE) model using the form of horizon entropy proposed by Viaggiu in 2014. This form of entropy is a generalization of the usual Bekenstein-Hawking entropy, having an extra term arising due to the dynamical nature of horizons in an expanding universe. We examine this new HDE model in the context of a flat Friedmann-Lema\^itre-Robertson-Walker universe filled with two cosmic fluids --- dark matter in the form of dust and holographic dark energy generated by Viaggiu entropy. We consider the Hubble horizon and the future event horizon as characteristic length scales and study the evolution of the Universe within these frameworks. Our analysis reveals some intriguing findings that include a possible cosmic doomsday scenario in the future, and certain observations are in striking contrast to other HDE models studied in the literature.
%PACS Numbers: 98.80.-k; 98.80.Cq\\\\

\end{abstract}

\maketitle

%%%%%%%%%%%%%%%%%%%%%%%%%%%%%%%%%%%%%%%%%%%%%%%%%%%%%%%%%%%%%%%%%%%%%%%%%%%%%%%%%%%%%%%%%%%%%%%%%%%%%%%%%%%%%%%%%%%%%%%%%%%%%%%%%%%%%%%%%%%%%%%%%%%%%%%%%%%%%%

\section{\label{sec-1} Introduction}
Dark energy (DE) has been one of the greatest mysteries in contemporary cosmology ever since its effects were measured by two independent research teams towards the end of the previous century using observational evidence \cite{Riess1,Perlmutter1}. Subsequent research has established the present standard model of cosmology to be the $\Lambda$CDM model in which $\Lambda$ refers to the DE candidate ("Cosmological Constant") responsible for the late-time accelerated expansion of the Universe, while CDM stands for "Cold Dark Matter" which is believed to have generated the large-scale structure of the Universe. Although observational data from modern state-of-the-art telescopes continue to favor the $\Lambda$CDM model, the nature and origin of DE is still beyond our ability to comprehend. Moreover, the $\Lambda$CDM model suffers from serious challenges \cite{Peebles1,Padmanabhan1,Perivolaropoulos1}. Numerous alternatives to this model have been proposed and studied in the literature in order to alleviate these problems, but none of them can be considered superior compared to $\Lambda$CDM when tested against observational data \cite{Copeland1,Frieman1,Caldwell1,Silvestri1,Li0,Bamba1}.\\

The holographic dark energy (HDE) paradigm is one such alternative which is based on the application of the holographic principle to the dark energy (DE) problem. This celebrated principle was first proposed by 't Hooft \cite{tHooft1} in 1993 and states that the information contained within a volume of space can be represented as a hologram, and this leads to a theory conceived on the boundary of that space. Two years later, Susskind \cite{Susskind1} provided an interpretation of the principle in the context of string theory. It is widely believed that the most successful realization of the holographic principle is the AdS/CFT correspondence, a conjecture which was put forward by Maldacena \cite{Maldacena1}. The first HDE model was proposed by Li \cite{Li1} in 2004. In this pioneering work, Li used the holographic principle to write the energy density of DE as proportional to the reduced Planck constant $M_p=1/\sqrt{8\pi G}$, where $G$ is Newton's gravitational constant, and the negative second power of the cosmological length scale which was chosen to be the future event horizon of the Universe. The model agreed well with the observational data available at that time, making the HDE model an acceptable alternative to the $\Lambda$CDM model. Furthermore, analysis by the CANTATA Network \cite{Akrami1} has revealed that holographic scenarios are not plagued by potential problems that are likely to appear in models of modified gravity. A host of HDE models have been obtained simply by modifying the form of the underlying entropy. Such extended forms include the nonadditive entropy due to Tsallis \cite{Tsallis1}, relativistic entropy due to Kaniadakis \cite{Kaniadakis1,Kaniadakis2}, power-law-corrected entropies \cite{Das1,Radicella1}, and quantum-gravity-corrected entropy due to Barrow \cite{Barrow1} among others. These extended entropy forms have, respectively, generated Tsallis HDE \cite{Saridakis0,Sadri1}, Kaniadakis HDE \cite{Drepanou1,Almada1}, power-law HDE \cite{Telali1}, and Barrow HDE \cite{Saridakis1}. One must note that these extended-entropy scenarios involve a constant parameter that quantifies the deviation from the standard Bekenstein-Hawking entropy. These models (and some less familiar HDE models such as Sharma-Mittal HDE \cite{Jahromi1} and R\'enyi HDE \cite{Moradpour1}) have also been studied in the context of different length scales and also with dark matter interactions. One may refer to the nice and extensive review by S. Wang, Y. Wang, and M. Li \cite{Wang1} and the references therein to get an overview of the HDE paradigm and its various extensions.\\

By using suitable theorems for the formation of black holes in a spatially flat, expanding Friedmann universe, Viaggiu \cite{Viaggiu1} managed to obtain a generalization of the Bekenstein-Hawking entropy. He later studied this generalized entropy in the context of dynamical apparent horizon in cosmology \cite{Viaggiu2}. In this regard, he obtained a generalized expression for the internal energy which turns out to be a constant of motion when measured at the dynamical apparent horizon of the Universe. This result is in strong agreement with the holographic principle. Interestingly, using gravitational thermodynamics, one of the authors has shown that \cite{Saha1} this extended Bekenstein-Hawking entropy forbids the phantom era, a result which is quite different from that obtained with the usual Bekenstein-Hawking entropy law and is in excellent agreement with current observational data.\\

In the present work, we investigate the possibility of a new HDE model considering the form of entropy obtained by Viaggiu \cite{Viaggiu1}. The paper is organized as follows. Section \ref{sec-2} gives a brief outline of the mathematical formulation of the generalized Bekenstein-Hawking entropy (hereafter we shall call it the Viaggiu entropy) worked out by Viaggiu \cite{Viaggiu1}. In Section \ref{sec-3}, we construct the HDE model based on Viaggiu entropy with the Hubble horizon as the characteristic length scale. Section \ref{sec-4} discusses the HDE model considering the characteristic length scale as the future event horizon and also studies its implications on the evolution of the Universe. Finally, Section \ref{sec-5} presents a short discussion and a summary of our work.

\section{\label{sec-2} A brief outline of the formulation of the Viaggiu entropy}
It is quite well-known that the Bekenstein-Hawking entropy was originally conjectured by Bekenstein \cite{Bekenstein0} and mathematically formulated a few years later by Hawking \cite{Hawking1} for a static spherically symmetric black hole in an asymptotically flat spacetime.\\

In his pioneering work, Viaggiu \cite{Viaggiu1,Viaggiu2} started with the condition that does not allow the formation of trapped surfaces \cite{Koc1,Brauer1,Malec1,Brauer2} in spacetimes that admit a mass concentration on spacelike surfaces with a pure trace extrinsic curvature. This is realized by the inequality
\begin{equation} \label{ineq-1}
\delta M \frac{G}{c^2} < \frac{L}{2} + \frac{AH}{4\pi c},  
\end{equation}
where $\delta M$ is the proper mass excess of the sphere $S$ (say) under consideration and $L$, $A$ are the proper radius and the area of the sphere respectively, $H$ is the Hubble parameter, $G$ is the Newton's gravitational constant, and $c$ is the speed of light in vacuum. Also, we have the universal upper bound for the entropy-to-energy ratio
\begin{equation} \label{ineq-2}
\frac{S}{E} \leqslant \frac{2\pi \kappa_B}{\hbar c}R, 
\end{equation}
obtained by Bekenstein \cite{Bekenstein1} for any arbitrary bounded system having an effective radius $R$. Here, $\kappa_B$ is the Boltzmann constant and $\hbar$ is the reduced Planck constant.\\

By using the Bekenstein bound (\ref{ineq-2}) together with inequality (\ref{ineq-1}), Viaggiu \cite{Viaggiu1} obtained the usual entropy law
\begin{equation} \label{bent}
S_{BH}=\frac{\kappa_B A}{4L_{p}^{2}}
\end{equation}
valid for static asymptotically flat spacetimes ($H=0$). In this case, $L$ is assumed to be the proper length of the event horizon of the black hole. Note here that $A$ is the proper area of the black hole and $L_p=\sqrt{\frac{\hbar G}{c^3}}$ is the Planck length. On the other hand, in a spatially flat, Friedmann expanding universe, one expects that the entropy should possess an (extra) work term which would appear as a manifestation of dynamic horizons ($H \neq 0$). To be precise, Viaggiu \cite{Viaggiu1} obtained the modified form
\begin{equation} \label{vent}
S_{BH}=\frac{\kappa_B A}{4L_{p}^{2}}+\frac{3\kappa_B}{2cL_{p}^{2}}VH,
\end{equation}
where $V$ is the effective geometric volume of the black hole measured at its outer apparent horizon. The standard Bekenstein-Hawking entropy is recovered when $H$ is set to zero in the above equation. It is worth mentioning that $L$ is considered to be the outer apparent horizon of the black hole \cite{Hayward1,Hayward2} in this case since it is a formidable task to identify its event horizon in an expanding universe.

\section{\label{sec-3} Viaggiu HDE with Hubble horizon as the IR cut-off}
We shall devote this section to the construction of a HDE model with the Viaggiu entropy given in Eq. (\ref{vent}). Hereinafter, we shall call this model the Viaggiu holographic dark energy (VHDE) model. It deserves to mention here that Cohen and collaborators \cite{Cohen1} were the first to propose a relationship between a short distance ultraviolet (UV) cutoff and a long distance infrared (IR) cutoff in local quantum field theory because of the limit set by the formation of a black hole. Indeed, if $\rho_{d}$ is the quantum zero-point energy density arising from a short-distance cutoff, then the total energy in a region of size $L$ should not exceed the mass of a black hole of equal size. Mathematically, one must have \cite{Li1,Bekenstein1,Wang1}
\begin{equation} \label{ineq}
M_{p}^{-2}L^4 \rho_{d} \leq S,
\end{equation}
where $L$ is the radius of the horizon and $S$ is the entropy. In terms of the length scale $L$, the Viaggiu entropy in Eq. (\ref{vent}) can be rewritten as\footnote{Henceforth, we use gravitational units where $G$, $c$, $\kappa_B$, $\hbar$ equal 1.}
\begin{equation}
S=\pi L^2 +2\pi H L^3
\end{equation}
since the area and the volume bounded by a horizon of radius $L$ are given by $4\pi L^2$ and $\frac{4}{3}\pi L^3$ respectively. Using this entropy and noting that the largest $L$ allowed is the one that saturates the inequality (\ref{ineq}), we arrive at
\begin{eqnarray} \label{rhodl}
\rho_d &=& \frac{\delta^2}{8\pi} L^{-4} S \nonumber \\
&=& \frac{\delta^2}{8\pi} L^{-4}(\pi L^2 +2\pi H L^3),
\end{eqnarray}
where $\delta^2$ is a positive numerical constant, not necessarily dimensionless. This parameter is introduced to account for neglected numerical factors in the last expression.\\

In this section, we shall identify the length scale $L$ with the Hubble horizon, so that $L=H^{-1}$. Then, from Eq. (\ref{rhodl}), we obtain the DE density as
\begin{equation} \label{rhod}
\rho_d=\frac{3}{8}\delta^2 H^2.
\end{equation}

\subsection{\label{sec-3.1} The Non-interacting Case}
For a flat FLRW universe filled with pressureless DM (dust) and DE in the form of VHDE, the Friedmann and acceleration equations are given by
\begin{equation} \label{feq}
3H^2=8\pi (\rho_m+\rho_d)
\end{equation}
and
\begin{equation} \label{aeq}
\dot{H}=-4\pi(\rho_m+p_m+\rho_d+p_d),
\end{equation}
where $\rho_m$ is the energy density of DM and $p_d$ is the pressure induced by VHDE. Since there is no interaction between the dark fluids, DM and DE will obey their individual energy-momentum conservation laws given by
\begin{equation} \label{ecl-m}
\dot{\rho}_m+3H(\rho_m+p_m)=0
\end{equation} 
and
\begin{equation} \label{ecl-d}
\dot{\rho}_d+3H(1+w_d)\rho_d=0
\end{equation}
respectively, where $p_m$ is the pressure generated by DM (which equals $0$ for dust) and $w_d=p_d/\rho_d$ is the equation of state (EoS) of DE. Now, since $p_m=0$, we must have
\begin{equation}
\rho_m=\rho_{m0}a^{-3}
\end{equation} 
from Eq. (\ref{ecl-m}), where $\rho_{m0}$ as the current energy density of DM and $a$ is the scale factor of the Universe. Substituting the expressions for $\rho_d$ and $\rho_m$ thus obtained, the Friedmann equation (\ref{feq}) yields the Hubble parameter $H$ in terms of the scale factor $a$ as
\begin{equation}
H(a)=\sqrt{\frac{8\pi \rho_{m0}}{3(1-\pi \delta^2)}} \times a^{-3/2}.
\end{equation}
The above equation can be integrated to obtain the time-dependent scale factor $a(t)$ as
\begin{equation}
a(t)=\left(\frac{3}{2}\right)^{2/3} \left\{\frac{8\pi \rho_{m0}}{3(1-\pi \delta^2)}\right\}^{1/3} t^{2/3}.
\end{equation}
Now, differentiating Eq. (\ref{rhod}) with respect to cosmic time $t$, we get $\dot{\rho}_d=\frac{3}{4}\delta^2 H\dot{H}$. Then, from Eq. (\ref{ecl-d}), we obtain 
\begin{equation} \label{wd-0}
w_d=0, 
\end{equation}
after a bit of simplification. Thus, the EoS of DE vanishes in this model. This implies that a DE-dominated phase is not possible with a noninteracting VHDE model if the Hubble horizon is considered as the IR cut-off. The effective EoS\footnote{$\Omega_d$ is the fractional energy density of VHDE, defined after Eq. (\ref{dr}).}, $w_{\mbox{eff}}=w_d \Omega_d$ also vanishes in this framework, which reduces the model to a dusty universe and is therefore uninteresting from the point of view of cosmology. It is interesting to note that this result is consistent with that obtained in the standard HDE model \cite{Li1}.

\subsection{\label{sec-3.2} The Interacting Case}
We are now forced to impose a component of interaction between DM and DE. In the literature, the interaction term is generally assumed to be proportional to the energy densities of DM and DE, which allows us to arrive at a scaling solution, thus alleviating the cosmic coincidence problem. Keeping this fact in mind, we assume an interaction of the form $Q=3b^2H(\rho_m+\rho_d)$ with $b^2$ as the coupling constant. This is a more general form of interaction than those in which the interaction terms are assumed to be proportional to either of the energy densities.\\

The interaction term will lead to a modification in the energy-momentum conservation equations for DM and DE in Eqs. (\ref{ecl-m}) and (\ref{ecl-d}) and the modified equations become
\begin{equation} \label{eclq-m}
\dot{\rho}_m+3H(\rho_m+p_m)=Q
\end{equation} 
and
\begin{equation} \label{eclq-d}
\dot{\rho}_d+3H(1+w_d)\rho_d=-Q
\end{equation}
respectively. If we now set $r=\rho_m/\rho_d$, the ratio of the energy densities, we obtain
\begin{equation} \label{dr}
\dot{r}=3Hw_dr+3H(1+r)^2
\end{equation}
using Eqs. (\ref{eclq-m}) and (\ref{eclq-d}). Also, using fractional energy densities $\Omega_m=8\pi\rho_m / 3H^2$ and $\Omega_d=8\pi\rho_d / 3H^2$ and the Friedmann equation $\Omega_m+\Omega_d=1$, we have $r=\frac{1}{\Omega_d}-1$ and $\dot{r}=-\dot{\Omega}_{d}/\Omega_{d}^{2}$. Equating this value with that obtained in Eq. (\ref{dr}), we obtain
\begin{equation} \label{wd}
w_d=-\frac{\Omega_{d}^{'}}{3\Omega_d(1-\Omega_d)}-\frac{b^2}{\Omega_d (1-\Omega_d)},
\end{equation}
where the prime denotes the derivative with respect to $x=\mbox{ln}~a$. It is interesting to note that the effect of the interaction term $Q$ is exhibited only through the second terms appearing in Eqs. (\ref{dr}) and (\ref{wd}). In other words, whatever the form of $Q$, the first terms of these two equations will remain unchanged.\\

It is surprising to see that if we take the derivative of the fractional DE density $\Omega_d=8\pi\rho_d / 3H^2$ with respect to $x$, we obtain $\Omega_{d}^{'}=0$. This implies that the fraction of the total energy density attributed to DE remains the same throughout the evolution of the Universe. To be precise, $\Omega_d=8\pi\rho_d / 3H^2=\pi \delta^2$, using Eq. (\ref{rhod}). A constant density of VHDE also implies a constant density of matter, and interestingly, this case hints at an Einstein static universe. This is a peculiar feature of this VHDE model, which is in striking contrast to other HDE models studied in the literature, with the Hubble horizon as the IR cutoff.\\ 

We are now compelled to consider the problem in the context of a different IR cut-off. However, at this point, we feel it necessary to outline the key characteristic length scales studied in the literature in the context of HDE models. It is important to note that the physics of any HDE model, and therefore the evolution of the Universe, depends on how this length scale is defined. The primary length scales considered in HDE cosmologies are (i) the Hubble horizon, $L=H^{-1}$, (ii) the particle horizon, $L=a(t)\int_{0}^{t} \frac{dt}{a(t)}$, (iii) the future event horizon, $L=a(t)\int_{t}^{\infty} \frac{dt}{a(t)}$, (iv) the Ricci scalar curvature, $L \propto R^{-\frac{1}{2}}$ \cite{Gao1}, and (v) the Granda-Oliveros cut-off, $L \propto (\alpha H^2+\beta \dot{H})^{-\frac{1}{2}}$ \cite{Granda1}. The Hubble horizon is the simplest choice for $L$ and, in principle, this choice solves the fine-tuning problem, nevertheless, the Hubble horizon leads to a vanishing EoS of DE in the standard HDE \cite{Li1} when no interaction terms are introduced. We have already shown that this result matches that obtained in the VHDE model. The interacting case also does not offer any remedy in our model. The particle horizon, on the other hand, fails to reproduce the observed late-time cosmic acceleration in the standard HDE \cite{Li1} because the EoS of DE does not admit values below $-\frac{1}{3}$ in this case. This drawback has led researchers to restrict its study in later HDE models. It is interesting to note that the most favored choice in the literature for the characteristic length scale in different HDE models has been the future event horizon because it explains the late-time cosmic acceleration and, most importantly, it fits the observational data well. In addition, these models allow the EoS of DE to cross the phantom divide in the future  \cite{Saridakis0,Sadri1,Drepanou1,Almada1,Telali1,Saridakis1,Jahromi1,Moradpour1}.\\

Taking a cue from the literature, in the next section, we study the VHDE model considering the future event horizon as the IR cut-off. We wish to see if our deductions keep in line with those obtained in other HDE models in this context.\footnote{We do not discuss the Ricci scalar curvature and Granda-Oliveros cut-offs in detail because they are not relevant to the present work.}

\section{\label{sec-4} Viaggiu HDE with the future event horizon as the IR cut-off}
It is well-known that the future event horizon exists only in an accelerating universe and is mathematically defined as the integral \cite{Faraoni1}
\begin{equation} \label{re-def}
R_E=a(t)\int_{t}^{\infty} \frac{dt}{a(t)}=a\int_{a}^{\infty} \frac{da}{Ha^2}=a\int_{x}^{\infty} \frac{dx}{Ha}.
\end{equation}
Physically speaking, the future event horizon is the proper distance to the most distant event that the comoving observer will ever see \cite{Faraoni1}.\\

Now, replacing $L$ with $R_E$ in Eq. (\ref{rhodl}), the DE density is recast as 
\begin{equation}\label{rhod-re}
\rho_d=\frac{\delta^2}{8R_{E}^2}(1+2HR_E).
\end{equation}
In this work, we only consider the case where DM and DE do not interact. Consequently, the EoS of DE becomes
\begin{equation} \label{wd-new}
w_d=-\frac{\Omega_{d}^{'}}{3\Omega_d(1-\Omega_d)},
\end{equation}
achieved by putting $b^2=0$ in Eq. (\ref{wd})\footnote{Note that Eq. (\ref{wd}) is true regardless of the choice of the characteristic length scale.}. It is apparent that we need to determine $\Omega_{d}^{'}$ in order to get a clear idea of $w_d$.\\

First, we calculate the time-derivative of the DE density:
\begin{equation}
\dot{\rho}_d=\frac{\delta^2}{4R_{E}^{3}}\left[1+R_{E}^{2}(\dot{H}-H^2)\right].
\end{equation}
Next, we evaluate the time-derivative of the fractional DE density:
\begin{equation}
\dot{\Omega}_{d} =\frac{2\pi \delta^2}{3H^2R_{E}^{2}}\left[-R_E(\dot{H}+H^2)+\frac{1}{R_E}-\frac{\dot{H}}{H}\right].
\end{equation}
Considering the derivative of $\Omega_{d}$ with respect to $\mbox{ln}~a$, the above equation modifies as
\begin{equation}\label{omd-dash}
\Omega_{d}^{'}=\frac{2\pi \delta^2}{3H^2R_{E}^{2}}\left[-R_E\frac{\dot{H}}{H}-HR_E+\frac{1}{HR_E}-\frac{\dot{H}}{H^2}\right].
\end{equation}
Rewriting Eq. (\ref{rhod-re}) in the form
\begin{equation}
\frac{3H^2\Omega_{d}}{8\pi}=\frac{\delta^2(1+2HR_E)}{8R_{E}^2},
\end{equation}
we arrive at a quadratic equation in $HR_E$:
\begin{equation} \label{quad-e}
(HR_{E})^2-\left(\frac{2 \pi \delta^2}{3\Omega_{d}}\right)HR_E-\frac{\pi \delta^2}{3\Omega_{d}}=0.
\end{equation}
Solving Eq. (\ref{quad-e}), we obtain
\begin{equation}
HR_E=\frac{\pi \delta^2}{3\Omega_d}\left(1+\sqrt{1+\frac{3\Omega_d}{\pi \delta^2}}\right), \frac{\pi \delta^2}{3\Omega_d}\left(1-\sqrt{1+\frac{3\Omega_d}{\pi \delta^2}}\right).
\end{equation}
We ignore the second root of $HR_E$ because it yields a negative value for $HR_E$, which is absurd. So, we must have
\begin{equation}\label{hre}
HR_E=\frac{\pi \delta^2}{3\Omega_d}\left(1+\sqrt{1+\frac{3\Omega_d}{\pi \delta^2}}\right).
\end{equation}
%Using Eqs. (\ref{quad-e}) and (\ref{hre}) in Eq. (\ref{omd-dash}), we obtain a more convenient form for $\Omega_{d}^{'}$:

%\onecolumngrid
%\begin{equation} \label{omd-dash-1}
%\Omega_{d}^{'}=\frac{2\Omega_d\left\{\frac{3}{2}(1+w_d\Omega_d)\sqrt{1+\frac{3\Omega_d}{\pi \delta^2}}\left(1+\sqrt{1+\frac{3\Omega_d}{\pi \delta^2}}\right)^2-2\left(1+\sqrt{1+\frac{3\Omega_d}{\pi \delta^2}}\right)+\frac{3\Omega_d}{\pi \delta^2}\left(\frac{3\Omega_d}{\pi \delta^2}-1\right)\right\}}{\left(1+\sqrt{1+\frac{3\Omega_d}{\pi \delta^2}}\right)\left[\frac{3\Omega_d}{\pi \delta^2}+2\left(1+\sqrt{1+\frac{3\Omega_d}{\pi \delta^2}}\right)\right]}
%\end{equation}

%\twocolumngrid
%Substituting this expression of $\Omega_{d}^{'}$ in Eq. (\ref{wd-new}), we obtain the EoS of DE:

%\onecolumngrid
%\begin{equation}
%w_d=-\frac{3\sqrt{1+\frac{3\Omega_d}{\pi \delta^2}}\left(1+\sqrt{1+\frac{3\Omega_d}{\pi \delta^2}}\right)^2-4\left(1+\sqrt{1+\frac{3\Omega_d}{\pi \delta^2}}\right)+\frac{6\Omega_d}{\pi \delta^2}\left(\frac{3\Omega_d}{\pi \delta^2}-1\right)}{3\left(1+\sqrt{1+\frac{3\Omega_d}{\pi \delta^2}}\right)\left[(1-\Omega_d)\left\{\frac{3\Omega_d}{\pi \delta^2}+2\left(1+\sqrt{1+\frac{3\Omega_d}{\pi \delta^2}}\right)\right\}+\Omega_d\sqrt{1+\frac{3\Omega_d}{\pi \delta^2}}\left(1+\sqrt{1+\frac{3\Omega_d}{\pi \delta^2}}\right)\right]}
%\end{equation}

%\twocolumngrid
%One can see that the expression for the EoS of DE has a complicated form and the only way to study its properties is by looking at its variation with respect to a more relevant cosmological parameter such as the scale factor $a$ or the redshift $z$. Now, 
\noindent
Using the above equation and Eq. (\ref{re-def}), we find that
\begin{equation}\label{re}
R_E=a\int_{x}^{\infty} \frac{dx}{Ha}=\frac{\pi \delta^2}{3H\Omega_d}\left(1+\sqrt{1+\frac{3\Omega_d}{\pi \delta^2}}\right).
\end{equation}
Again, from the Friedmann equation, one obtains
\begin{equation}
\frac{1}{Ha}=\frac{\sqrt{a(1-\Omega_d)}}{H_0 \sqrt{\Omega_{m0}}},
\end{equation}
where $\Omega_{m0}$ is the current value of the fractional dark matter density. Taking the derivative with respect to the parameter $x$ on both sides, we arrive at
\begin{equation} \label{ddx}
\frac{d}{dx}\left(\frac{1}{Ha}\right)=\frac{a(1-\Omega_{d}-\Omega_{d}^{'})}{2H_{0}\sqrt{a\Omega_{m0}(1-\Omega_{d})}}.
\end{equation}
Finally, we take the derivative with respect to $x$ on both sides of the last equality in Eq. (\ref{re}) and use Eq. (\ref{ddx}) to obtain

\onecolumngrid
\begin{eqnarray} \label{omdd}
\frac{\Omega_{d}^{'}}{\Omega_{d}(1-\Omega_{d})}=\frac{\frac{6\Omega_d}{\pi \delta^2}+\left(1+\sqrt{1+\frac{3\Omega_d}{\pi \delta^2}}\right)}{(2-\Omega_d)\left(1+\sqrt{1+\frac{3\Omega_d}{\pi \delta^2}}\right)-\frac{\frac{3\Omega_d}{\pi \delta^2}(1-\Omega_d)}{\sqrt{1+\frac{3\Omega_d}{\pi \delta^2}}}}.
\end{eqnarray}

\begin{figure*}[ht]
     \centering
     \begin{minipage}{0.325\textwidth}
         \centering
         \includegraphics[width=\textwidth]{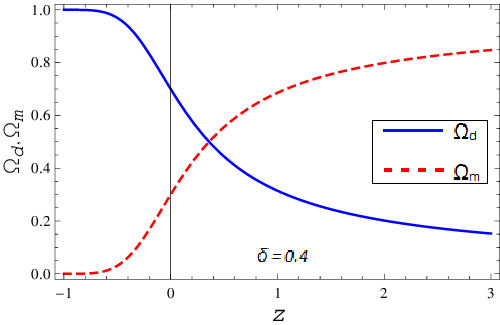}
%        \caption{}
%        \label{}
     \end{minipage}
     \hfill
     \begin{minipage}{0.325\textwidth}
         \centering
         \includegraphics[width=\textwidth]{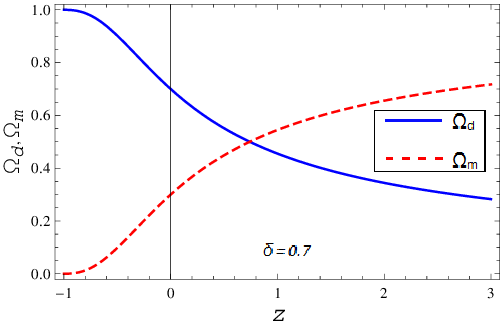}
%        \caption{}
%        \label{}
     \end{minipage}
     \hfill
     \begin{minipage}{0.325\textwidth}
         \centering
         \includegraphics[width=\textwidth]{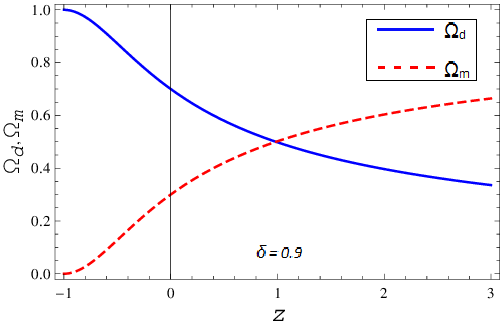}
%        \caption{}
%        \label{}
     \end{minipage}
        \caption{The evolution of the density parameters of DM and VHDE as a function of the redshift $z$ corresponding to $\delta=0.4$ (left panel), $\delta=0.7$ (middle panel), and $\delta=0.9$ (right panel). We have imposed the initial condition $\Omega_d(x=-\mbox{ln}(1+z)=0) \equiv \Omega_{d0} \approx 0.7$.}
        \label{fig:omm-omd}
\end{figure*}

\twocolumngrid
\noindent
The above equation describes the evolution of the density parameter of VHDE, but it is a challenging task to obtain an analytical solution of this equation. Fortunately, it can be solved numerically. First, we note that $x \equiv \mbox{ln}~a=-\mbox{ln}(1+z)$ so that $\Omega_{d}^{'}$ transforms to $-(1+z)\frac{d\Omega_d}{dz}$. Then, imposing the initial condition $\Omega_d(x=-\mbox{ln}(1+z)=0) \equiv \Omega_{d0} \approx 0.7$, in accordance with the observational data \cite{Ade1}, we obtain a numerical solution\footnote{The numerical solution and all the plots have been generated with the help of the MATHEMATICA software.} of Eq. (\ref{omdd}). Figure \ref{fig:omm-omd} shows the variation of $\Omega_m(z)$ and $\Omega_d(z)=1-\Omega_m(z)$, respectively, the density parameters corresponding to DM and VHDE, as a function of the redshift $z$. The three panels, from left to right, correspond, respectively, to the three choices of the parameter $\delta$, viz. $\delta=0.4$, $\delta=0.7$, and $\delta=0.9$. It is evident from the figures that the VHDE, like other HDE models, alleviates the cosmic-coincidence problem. Furthermore, we observe that the VHDE starts to dominate much earlier when the value of the parameter $\delta$ is higher. Such a scenario would suppress structure formation in the Universe as gravity would not be able to clump matter in an efficient matter, thus leading to fewer large galaxies and galaxy clusters, resulting in a distribution of cosmic structure that is quite different from what is observed today \cite{Bartelmann1,Grossi1,Pu1,Sobotka1}. Early dark energy (EDE) models have been studied extensively in the literature, for a review, one may see Ref. \cite{Poulin1} and references [79-111] therein. Although the concept of EDE has been around since the dawn of the 21st century \cite{Doran1,Wetterich1,Doran2}, these models have gained renewed attention in the last decade as a possible way to resolve the Hubble tension \cite{Karwal1,Poulin2,Shen1}.  Another thing that grabs our attention in Eq. (\ref{omdd}) is that $\Omega_{d}^{'}$ is always nonnegative, as one should expect. This can be deduced from the fact that the sign of $\Omega_{d}^{'}$ is determined solely by the sign of the denominator on the right hand side of the equation, which is always positive for $\Omega_{d} \in [0,1]$. This implies that the Universe will continue to expand forever.\\

In order to get an idea about the density of VHDE in the early universe, following Ref. \cite{Li1}, we let $\epsilon=\Omega_d$ and recast Eq. (\ref{omdd}) in the form
\begin{equation} \label{deps}
\epsilon'=\frac{d\epsilon}{dx}=\frac{2\epsilon + \left(\frac{21}{2\pi \delta^2}-2\right)\epsilon^2}{4+(-2+\frac{6}{\pi \delta^2})\epsilon+\left(-\frac{9}{4\pi^2 \delta^4}-\frac{3}{2\pi \delta^2}\right)\epsilon^2},    
\end{equation}
where we have used binomial expansions whenever applicable and kept terms upto the second order in $\epsilon$. By writing partial fractions, the above equation can be solved exactly. The solution can be written in terms of the scale factor $a$ as
\begin{equation}
2~\mbox{ln}\left(\frac{\epsilon}{h\epsilon+2}\right)+\left(\frac{f}{h}-\frac{2g}{h^2}\right)~\mbox{ln}(h\epsilon+2)+\frac{g\epsilon}{h}=\mbox{ln}~a+x_0,    
\end{equation}
where $f=-2+\frac{6}{\pi \delta^2}$, $g=-\frac{9}{4\pi^2 \delta^4}-\frac{3}{2\pi \delta^2}$, $h=\frac{21}{2\pi \delta^2}-2$, and $x_0$ is the integration constant that can be determined at the present epoch by letting $a=1$.\\\\
Solving (\ref{deps}), we arrive at
\begin{equation}
\epsilon = \Omega_{d} = 2^{\frac{2234}{3481}}\mbox{e}^{\frac{x_0}{2}}\sqrt{a}
\end{equation}
and thus
\begin{equation}
\rho_d \approx \Omega_d \rho_m = \Omega_d \rho_{m0}a^{-3} = \rho_{m0} 2^{\frac{2234}{3481}} \mbox{e}^{\frac{x_0}{2}} a^{-\frac{5}{2}}.   
\end{equation}
It is observed that although the DE is higher for small values of $a$, it is still dominated by DM and the standard big bang theory holds firmly. 

\onecolumngrid

\begin{figure*}[ht]
     \centering
     \begin{minipage}{0.425\textwidth}
         \centering
         \includegraphics[width=\textwidth]{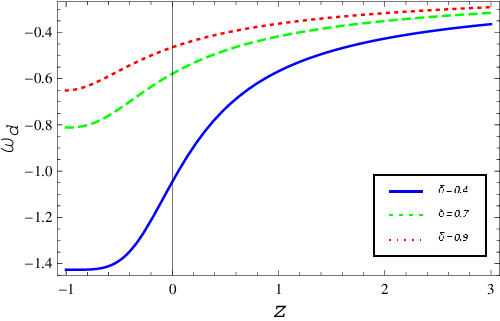}
%        \caption{}
%        \label{}
     \end{minipage}
     \hspace{1cm}
     \begin{minipage}{0.425\textwidth}
         \centering
         \includegraphics[width=\textwidth]{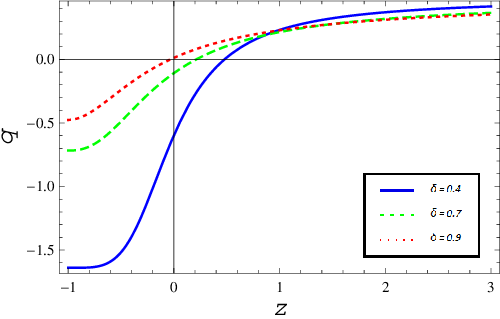}
%        \caption{}
%        \label{}
     \end{minipage}
        \caption{The left and the right panels, respectively, show the evolution of the EoS of VHDE and the deceleration parameter as a function of the redshift $z$ for the three choices of $\delta$, viz. $\delta=0.4$, $\delta=0.7$, and $\delta=0.9$.}
        \label{fig:eos-q}
\end{figure*}

\twocolumngrid
Now, on multiplying Eq. (\ref{omdd}) by $-\frac{1}{3}$, one obtains the EoS of VHDE as

\onecolumngrid
\begin{eqnarray} \label{wd-f}
w_d=-\frac{\Omega_{d}^{'}}{3\Omega_{d}(1-\Omega_{d})}=-\frac{1}{3}\left[\frac{\frac{6\Omega_d}{\pi \delta^2}+\left(1+\sqrt{1+\frac{3\Omega_d}{\pi \delta^2}}\right)}{(2-\Omega_d)\left(1+\sqrt{1+\frac{3\Omega_d}{\pi \delta^2}}\right)-\frac{\frac{3\Omega_d}{\pi \delta^2}(1-\Omega_d)}{\sqrt{1+\frac{3\Omega_d}{\pi \delta^2}}}}\right].
\end{eqnarray}

%\twocolumngrid
In addition, the deceleration parameter $q$ is obtained as
\onecolumngrid
\begin{eqnarray}
q &=& \frac{1}{2}(1+3w_d \Omega_d) \nonumber \\
&=& \frac{1}{2}-\frac{\Omega_d \sqrt{1+\frac{3\Omega_d}{\pi \delta^2}} \left[\left(1+\frac{3\Omega_d}{\pi \delta^2}\right)^{\frac{3}{2}}+\left(1+\frac{9\Omega_d}{2\pi \delta^2}\right)\right]}{\left(1+\sqrt{1+\frac{3\Omega_d}{\pi \delta^2}}\right)\left[\frac{3\Omega_d}{\pi \delta^2}+(2-\Omega_d)\left(1+\sqrt{1+\frac{3\Omega_d}{\pi \delta^2}}\right)\right]}
\end{eqnarray}

\twocolumngrid
In Figure \ref{fig:eos-q}, we show the evolution of the EoS $w_d$ and the deceleration parameter $q$ as a function of the redshift $z$ in the left and right panels, respectively. An important observation from these figures is that recent observational data prefer lower values of the parameter $\delta$. In fact, the values $\Omega_d=0.7$ and $w_d=-1$ yield $\delta \approx 0.42$ at the present epoch. However, it must be noted that the aforementioned values are not necessary to have a best fit with current data, rather a statistical fit with recent observational data is required. Furthermore, it is interesting to observe that the phantom era can be realized in this HDE scenario, and this may lead the Universe to end in a Big Rip \cite{Caldwell1}. However, the Big Rip scenario can be avoided with suitable values of $\delta$. In fact, the condition $w_d \geq -1$ yields the lower bound on $\delta^2$ 
\begin{equation}
\delta^2 \geq \frac{6\Omega_d}{\pi [(5-3\Omega_d)+2\sqrt{10-6\Omega_d}]},
\end{equation}
after a bit of algebra.

\section{\label{sec-5} Discussion and summary of the results}
This paper dealt with the formulation of a new HDE model, the Viaggiu holographic dark energy (VHDE), considering the form of entropy proposed by Viaggiu \cite{Viaggiu1} as a generalization of the usual Bekenstein-Hawking entropy. This generalized entropy takes into account the dynamical nature of horizons in an expanding universe. We investigated the features of the VHDE model in the context of a flat, FLRW universe filled with dark matter (DM) in the form of dust and VHDE. Initially, we considered the Hubble horizon as the IR cutoff, and found that the EoS of VHDE vanishes when there is no component of interaction between DM and VHDE. Interestingly, it was found that the density parameter corresponding to VHDE is a constant throughout the evolutionary history of the Universe, whether interaction between DM and VHDE is introduced or not. This result is consistent with an Einstein static universe. As far as our knowledge is concerned, these observations are in striking contrast to other HDE models with the Hubble horizon as the IR cutoff studied in the literature. Moreover, none of these results is consistent with the observational data. So, we then considered the future event horizon as the IR cutoff and devoted our attention to the noninteracting case only due to the apparent complexity in the equations. We obtained an evolution equation for the VHDE density parameter $\Omega_d$, which was then solved numerically using MATHEMATICA by imposing a suitable initial condition consistent with the latest observational data. We then studied the variations of $\Omega_d$ and $\Omega_m$ as a function of the redshift for three choices of the parameter $\delta$. The figures reveal that VHDE begins to dominate much earlier when the value of $\delta$ is higher. Also, this noninteracting VHDE model with the future event horizon as the IR cutoff alleviates the cosmic coincidence problem. We further deduce that the Universe will be ever-expanding and free from any future singularity. Finally, we obtained expressions for the EoS of VHDE and the deceleration parameter of the model. and studied their variations as a function of the redshift for the three choices of the parameter $\delta$. The figures show that lower values of $\delta$ are favored by the observational data. A simple calculation yields $\delta \approx 0.42$ at the present time. Another interesting observation is that the VHDE can behave as a phantom fluid in the far future, and the Universe may end in a cosmic doomsday, otherwise known as the Big Rip. Previously, Saridakis \cite{Saridakis1} has shown that Barrow HDE with the future event horizon as the IR cut off can explain the thermal history of the Universe, including a possible phantom-divide crossing scenario in the future. Recently, Sudharani et al. \cite{Sudharani1} have studied Barrow HDE in both non-interacting and interacting scenarios with the future event horizon as the IR cut-off and tested their models against observational data. Their analysis also confirms the present status of the Universe as an accelerating one. We observe that, when the future event horizon is assumed to be the IR cut-off, the evolutionary features of the VHDE model agree well with the findings of recent HDE models studied with a similar cut-off scale.\\

Future studies may investigate the effects of cosmological perturbations on our model and also undertake a detailed Bayesian analysis to distinguish it from the standard $\Lambda$CDM model. These analyzes will help to dig deeper into the intricate structure of the VHDE, and also identify any unique features compared to other HDE models in the literature.             

%%%%%%%%%%%%%%%%%%%%%%%%%%%%%%%%%%%%%%%%%%%%%%%%%%%%%%%%%%%%%%%%%%%%%%%%%%%%%%%%%%%%%%%%%%%%%%%%%%%%%%%%%%%%%

\begin{acknowledgments}
The first author is grateful to {\bf Sree Chaitanya College, Habra} and {\bf Jadavpur University} for extending all necessary administrative support during his Ph.D. research work. The second author wishes to thank the {\bf Inter-University Centre for Astronomy and Astrophysics (IUCAA), Pune} for their kind hospitality as a significant portion of this work was completed during a visit there in October 2024. All the authors thank {\bf Professor Subenoy Chakraborty} for his insightful comments on the first draft of this paper, which significantly improved the quality of the manuscript, and the anonymous peer-reviewer for a careful reading of the manuscript and providing many fruitful suggestions for its improvement.
\end{acknowledgments}

%%%%%%%%%%%%%%%%%%%%%%%%%%%%%%%%%%%%%%%%%%%%%%%%%%%%%%%%%%%%%%%%%%%%%%%%%%%%%%%%%%%%%%%%%%%%%%%%%%%%%%%%%%%%%
\vspace{0.25cm}
{\bf Data Availability Statement:} Data sharing is not applicable to this article as no datasets were generated or analyzed during the current study.-- 

%%%%%%%%%%%%%%%%%%%%%%%%%%%%%%%%%%%%%%%%%%%%%%%%%%%%%%%%%%%%%%%%%%%%%%%%%%%%%%%%%%%%%%%%%%%%%%%%%%%%%%%%%%%%%


\frenchspacing
\begin{thebibliography}{100}

\bibitem{Riess1} A. G. Riess {\it et al.}, {\color{magenta} Astron. J. {\bf 116}, 1009 (1998)}.
\bibitem{Perlmutter1} S. Permutter {\it et al.}, {\color{magenta} Astrophys. J. {\bf 517}, 565 (1999)}.
\bibitem{Peebles1} P. J. E. Peebles and B. Ratra, {\color{magenta} Rev. Mod. Phys. {\bf 75}, 559 (2003)}.
\bibitem{Padmanabhan1} T. Padmanabhan, {\color{magenta} Phys. Rep. {\bf 380}, 235 (2003)}.
\bibitem{Perivolaropoulos1} L. Perivolaropoulos and F. Skara, {\color{magenta} New Astron. Rev. {\bf 95}, 101659 (2022)}.
\bibitem{Copeland1} E. J. Copeland, M. Sami, and S. Tsujikawa, {\color{magenta} Int. J. Mod. Phys. D {\bf 15}, 1753 (2006)}.
\bibitem{Frieman1} J. Frieman, M. Turner, and D. Huterer, {\color{magenta} Ann. Rev. Astron. Astrophys. {\bf 46}, 385 (2008)}.
\bibitem{Caldwell1} R. R. Caldwell and M. Kamionkowski, {\color{magenta} Ann. Rev. Nucl. Part. Sci. {\bf 59}, 397 (2009)}.
\bibitem{Silvestri1} A. Silvestri and M. Trodden, {\color{magenta} Rep. Prog. Phys. {\bf 72}, 096901 (2009)}.
\bibitem{Li0} M. Li, X. -D. Li, S. Wang, and Y. Wang, {\color{magenta} Commun. Theor. Phys. {\bf 56}, 325 (2011)}.
\bibitem{Bamba1} K. Bamba, S. Capozziello, S. Nojiri, and S. D. Odintsov, {\color{magenta} Astrophys. Space Sci. {\bf 342}, 155 (2012)}.
\bibitem{tHooft1} G. 't Hooft, {\color{blue} arXiv: gr-qc/9310026}.
\bibitem{Susskind1} L. Susskind, {\color{magenta} J. Math. Phys. {\bf 36}, 6377 (1995)}.
\bibitem{Maldacena1} J. M. Maldacena, {\color{magenta} Int. J. Theor. Phys. {\bf 38}, 1113 (1999)}.
\bibitem{Li1} M. Li, {\color{magenta} Phys. Lett. B {\bf 603}, 1 (2004)}.
\bibitem{Akrami1} Y. Akrami {\it et al.}, {\it \color{blue} Modified Gravity and Cosmology: An Update by the CANTATA Network} (Springer, 2021).
\bibitem{Tsallis1} C. Tsallis, {\color{magenta} J. Stat. Phys. {\bf 52}, 479 (1988)}.
\bibitem{Kaniadakis1} G. Kaniadakis, {\color{magenta} Phys. Rev. E {\bf 66}, 056125 (2002)}.
\bibitem{Kaniadakis2} G. Kaniadakis, {\color{magenta} Phys. Rev. E {\bf 72}, 036108 (2005)}.
\bibitem{Das1} S. Das, S. Shankarnarayanan, and S. Sur, {\color{magenta} Phys. Rev. D {\bf 77}, 064013 (2008)}.
\bibitem{Radicella1} N. Radicella and D. Pav\'on, {\color{magenta} Phys. Lett. B {\bf 691}, 121 (2010)}.
\bibitem{Barrow1} J. D. Barrow, {\color{magenta} Phys. Lett. B {\bf 808}, 135643 (2020)}.
\bibitem{Saridakis0} E. N. Saridakis, K. Bamba, R. Myrzakulov and F. K. Anagnostopoulos, {\color{magenta} J. Cosmol. Astropart. Phys. {\bf 12}, 012 (2018)}.
\bibitem{Sadri1} E. Sadri, {\color{magenta} Eur. Phys. J. C {\bf 79}, 762 (2019)}.
\bibitem{Drepanou1}  N. Drepanou, A. Lymperis, E. N. Saridakis, and K. Yesmakhanova, {\color{magenta} Eur. Phys. J. C {\bf 82}, 449 (2022)}.
\bibitem{Almada1}  A. Hern\'andez-Almada {\it et al.}, {\color{magenta} Mon. Not. Roy. Astron. Soc. {\bf 511}, 4147 (2022)}.
\bibitem{Telali1} E. C. Telali and E. N. Saridakis, {\color{magenta} Eur. Phys. J. C {\bf 82}, 466 (2022)}.
\bibitem{Saridakis1} E. N. Saridakis, {\color{magenta} Phys. Rev. D {\bf 102}, 123525 (2020)}.
\bibitem{Jahromi1} A. Sahayan Jahromi {\it et al.}, {\color{magenta} Phys. Lett. B {\bf 780}, 21 (2018)}.
\bibitem{Moradpour1} H. Moradpour {\it et al.}, {\color{magenta} Eur. Phys. J. C {\bf 78}, 829 (2018)}.
\bibitem{Wang1} S. Wang, Y. Wang, and M. Li, {\color{magenta} Phys. Rep. {\bf 696}, 1 (2017)}.
\bibitem{Viaggiu1} S. Viaggiu, {\color{magenta} Mod. Phys. Lett. A {\bf 29}, 1450091 (2014)}.
\bibitem{Viaggiu2} S. Viaggiu, {\color{magenta} Gen. Relativ. Gravit. {\bf 47}, 86 (2015)}.
\bibitem{Saha1} S. Saha, {\color{magenta} Int. J. Mod. Phys. A {\bf 34}, 1950193 (2019)}.
\bibitem{Bekenstein0} J. D. Bekenstein, {\color{magenta} Lett. Nuovo Cimento {\bf 4}, 737 (1972)}.
\bibitem{Hawking1} S. W. Hawking, {\color{magenta} Commun. Math. Phys. {\bf 43}, 199 (1975)}.
\bibitem{Koc1} P. Koc and E. Malek, {\color{magenta} Acta Phys. Pol. B {\bf 23}, 123 (1992)}.
\bibitem{Brauer1} U. Brauer and E. Malec, {\color{magenta} Phys. Rev. D {\bf 45}, R1836 (1992)}.
\bibitem{Malec1} E. Malec and N. O. Murchadha, {\color{magenta} Phys. Rev. D {\bf 47}, 1454 (1993)}.
\bibitem{Brauer2} U. Brauer, E. Malec, and N. O. Murchadha, {\color{magenta} Phys. Rev. D {\bf 49}, 5601 (1994)}.
\bibitem{Bekenstein1} J. D. Bekenstein, {\color{magenta} Phys. Rev. D {\bf 23}, 287 (1981)}.
\bibitem{Hayward1} S. A. Hayward, {\color{magenta} Phys. Rev. D {\bf 49}, 6467 (1994)}.
\bibitem{Hayward2} S. A. Hayward, {\color{magenta} Class. Quantum Grav. {\bf 11}, 3025 (1994)}.
\bibitem{Cohen1} A. Cohen, D. Kaplan, and A. Nelson, {\color{magenta} Phys. Rev. Lett. {\bf 82}, 4971 (1999)}.
\bibitem{Gao1} C. Gao, F. Wu, X. Chen, and Y. -G. Shen, {\color{magenta} Phys. Rev. D {\bf 79}, 043511 (2009)}.
\bibitem{Granda1} L. N. Granda and A. Oliveros, {\color{magenta} Phys. Lett. B {\bf 671}, 199 (2009)}.
\bibitem{Faraoni1} V. Faraoni, {\it \color{blue} Cosmological and black hole apparent horizons} (Springer, 2015).
\bibitem{Ade1} P. A. R. Ade {\it et al.}, {\color{magenta} Astron. Astrophys. {\bf 594}, A13 (2016)}.
\bibitem{Bartelmann1} S. A. Hayward, {\color{magenta} Astron. Astrophys. {\bf 454}, 27 (2006)}.
\bibitem{Grossi1} M. Grossi and M. Springel, {\color{magenta} Mon. Not. R. Astron. Soc. {\bf 394}, 1559 (2009)}.
\bibitem{Pu1} B. -Y. Pu, X. -D. Xu, and B. Wang, {\color{magenta} Phys. Rev. D {\bf 92}, 123527 (2015)}.
\bibitem{Sobotka1} A. C. Sobotka, A. L. Erickcek, and T. L. Smith, {\color{magenta} Phys. Rev. D {\bf 111}, 123522 (2025)}.
\bibitem{Poulin1} V. Poulin, T. L. Smith, and T. Karwal, {\color{magenta} Phys. Dark Univ. {\bf 42}, 101348 (2023)}.
\bibitem{Doran1} M. Doran, M. J. Lilley, J. Schwindt, and C. Wetterich, {\color{magenta} Astrophys. J. {\bf 559}, 501 (2001)}.
\bibitem{Wetterich1} C. Wetterich, {\color{magenta} Phys. Lett. B {\bf 594}, 17 (2004)}.
\bibitem{Doran2} M. Doran and G. Robbers, {\color{magenta} J. Cosmol. Astropart. Phys. {\bf 06}, 026 (2006)}.
\bibitem{Karwal1} T. Karwal and M. Kamionkowski, {\color{magenta} Phys. Rev. D. {\bf 94}, 103523 (2016)}.
\bibitem{Poulin2} V. Poulin, T. L. Smith, T. Karwal, and M. Kamionkowski, {\color{magenta} Phys. Rev. Lett. {\bf 122}, 221301 (2019)}.
\bibitem{Shen1} X. Shen, M. Vogelsberger, M. Boylan-Kolchin, S. Tacchella, and R. P. Naidu, {\color{magenta} Mon. Not. R. Astron. Soc. {\bf 533}, 3923 (2024)}.
\bibitem{Caldwell2} R. R. Caldwell, M. Kamionkowski, and N. N. Weinberg, {\color{magenta} Phys. Rev. Lett. {\bf 91}, 071301 (2003)}.
\bibitem{Sudharani1} L. Sudharani, N. S. Kavya, and V. Venkatesha, {\color{magenta} Nucl. Phys. B {\bf 1009}, 116725 (2024)}.


%\bibitem{Wald1} R. M. Wald, {\color{magenta} Living Rev. Relativ. {\bf 4}, 6 (2001)}.
%\bibitem{Poisson1} E. Poisson, {\it A relativist’s toolkit: The mathematics of black hole mechanics} (Cambridge University Press, 2004).
%\bibitem{Helou1} A. Helou, arXiv: 1502.04235 [gr-qc].



\end{thebibliography}
\end{document}